\documentclass[twocolumn,aps,prl,showpacs,amsmath,amssymb,floatfix]{revtex4-1}
\usepackage{graphicx}
\usepackage{dcolumn}
\usepackage{bm}
\usepackage{xcolor}

\begin{document}

\title{Precursors to Molecular Slip on Smooth Hydrophobic Surfaces}

\author{Justin E. Pye}
\author{Clay E. Wood}
\author{Justin C. Burton}
\email{justin.c.burton@emory.edu}

\affiliation{Department of Physics, Emory University, Atlanta, Georgia 30322, USA}

\date{\today}

\begin{abstract}
Experiments and simulations suggest that simple liquids may experience slip while flowing near a smooth, hydrophobic surface. Here we show how precursors to molecular slip can be observed in the complex response of a liquid to oscillatory shear. We measure both the change in frequency and bandwidth of a quartz crystal microbalance (QCM) during the growth of a single drop of water immersed in an ambient liquid. By varying the hydrophobicity of the surface using self-assembled monolayers, our results show little or no slip for water on all surfaces. However, we observe excess transverse motion near hydrophobic surfaces due to weak binding in the corrugated surface potential, an essential precursor to slip. We also show how this effect can be easily missed in simulations utilizing finite-ranged interaction potentials.
\end{abstract}

\maketitle

The conventional ``no-slip" boundary condition for liquid flow near a solid wall assumes that the velocity is continuous. A more general boundary condition, first introduced by Navier in 1823, allows for a discontinuity in the tangential velocity, 
$\Delta v_t=b(\partial v_t/\partial n)$, where $n$ is the normal direction pointing into the liquid, and $b$ is the ``slip length"  \cite{Lauga2007}. Although negligible on the macroscale, slip is especially important for nanoscale transport in confined \cite{Squires2005,Bocquet2010} and biological flows \cite{Rasaiah2008}, and for technological processes involving nanofiltration \cite{Nair2012}, desalination \cite{Park2014}, and energy harvesting \cite{Joshi2014}. Highly structured, superhydrophobic surfaces can lead to effective slip lengths greater than 1 $\mu$m due to residual gas bubbles trapped in surface features. However, on smooth surfaces, reported values of $b$ range from 0 to 100 nm or more, and are sensitive to issues such as nano-confinement \cite{Bocquet1994,Ortiz2013,Secchi2016} and surface contamination or roughness \cite{Granick2003,Cottin-Bizonne2005,Lee2014,Vinogradova2006,Priezjev2006,Vinogradova2011}. 

Two parameters are thought to determine slip in Newtonian liquids on molecularly smooth surfaces, yet there is little agreement between theory and experiments. First, surface hydrophobicity, often measured by the equilibrium contact angle in air, enhances slip through weak surface interactions. However, for water on hydrophobic surfaces, slip lengths reported in simulations \cite{Huang2008,Ho2011} are systematically lower than those reported in experiments \cite{Neto2005,Joly2006,Bouzigues2008,McBride2009,Cottin-Bizonne2005,Vinogradova2009}. Second, a large shear rate at the liquid-solid interface will induce slip if the resulting shear stress is sufficiently strong \cite{Thompson1997,Bocquet2007,Martini2008}. However, experiments often operate at shear rates of $\dot{\gamma}\leq10^5$ s$^{-1}$, whereas simulations exceed $\dot{\gamma}=10^9$ s$^{-1}$, making direct quantitative comparisons ambiguous. 

Here we directly connect theory and experiments concerning molecular slip for water by systematically varying the surface hydrophobicity at large experimental shear rates. We find little or no slip ($|b|\leq$ 3 nm) on all surfaces for $\theta^a_{w}\leq120^{\circ}$, where $\theta^a_{w}$ is the advancing contact angle of a water drop in air. Before slip can occur near a surface, the liquid molecules must experience some average elastic displacement in local surface potential wells. We are able to measure these ``precursors" to slip, and show that the amplitude of displacement increases with surface hydrophobicity, providing a direct connection between the nanometer-scale liquid dynamics and surface interactions. In addition, our results allow us to predict when slip should occur at higher shear rates, and estimate the depletion length between the liquid and solid. Finally, we show how numerical techniques which are universally employed to simulate nanoscale flows, such as a cut-off radius for molecular interactions, can strongly influence measured slip lengths for hydrophobic surfaces.

\begin{figure*}
\includegraphics[width=6.3 in]{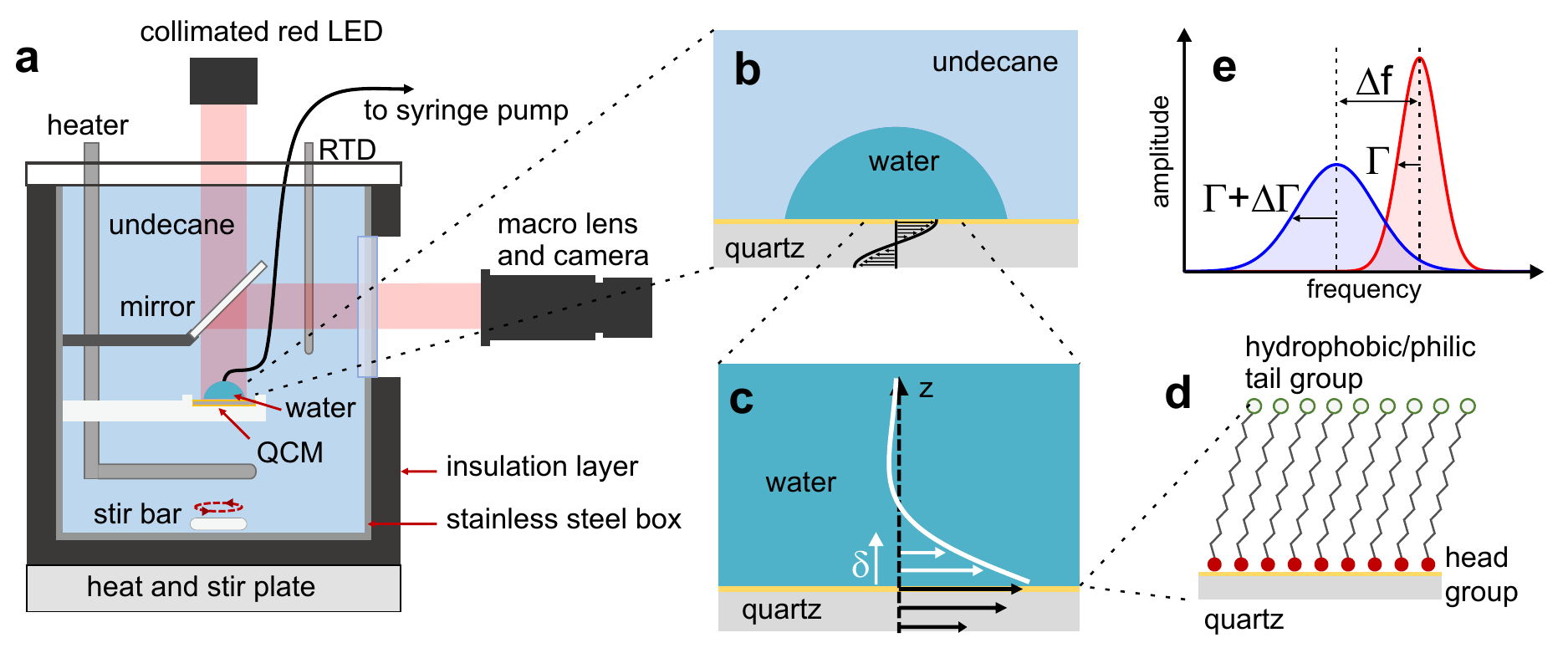}
\caption[]{(a) Diagram of the experimental setup. The QCM is immersed in a temperature-controlled bath of undecane as a water drop is slowly grown and imaged on the surface. (b-c) The QCM emits a decaying shear wave which penetrates a characteristic distance $\delta\approx 200$ nm into the liquid. (d) The hydrophobicity of the QCM surface is modified by oxygen plasma cleaning and/or coating with self-assembled monolayers. (e) The liquid's interaction with the surface adds inertia and damping to the crystal's motion, which decreases the resonant frequency by $\Delta f$ and increases the half-bandwidth by $\Delta \Gamma$.  
\label{fig_1}} 
\end{figure*}

Our experiments utilize a QCM exposed to an unconfined liquid bath. The maximum shear rate is $\dot{\gamma}=7.7\times10^5$ s$^{-1}$ \cite{supp}, which is nearly an order of magnitude larger than other experimental techniques used to measure slip. QCMs have been previously used to measure liquid-solid boundary conditions \cite{Ellis2003a,McHale2004,Du2004,Ellis2003b,Ferrante1994,Qiao2016} and are known to be sensitive to temperature and elastic stress \cite{Johannsmann2015}. To achieve sub-nanometer resolution, we implemented a new, pseudo-differential method which compares the boundary condition between a wetting liquid (undecane), and a non-wetting liquid (water). Figure \ref{fig_1} shows our experimental setup. An AT-cut QCM operating at its fundamental frequency ($f_0$ = 5 MHz) is immersed in a temperature-controlled bath of undecane. After a few hours of equilibration, a drop of water is slowly grown on the surface consisting of bare gold, gold coated with SiO$_2$, or gold coated with various self-assembled monolayers (SAMs) using a thiol-based chemistry \cite{supp}. All liquids were degassed prior to the experiment in order to minimize the effects of surface nanobubbles \cite{Lohse2015}. 

Upon immersion into a liquid with a no-slip boundary condition, the change in the resonant frequency and half-bandwidth of the QCM (Fig.\ \ref{fig_1}e) are given by \cite{Kanazawa1985,Johannsmann2015,supp}
\begin{align}
\label{shift1}
\Delta f_{liq}=-\Delta \Gamma_{liq}=& -\dfrac{f_0^{3/2}}{\pi Z_q}\left(\pi \eta \rho\right)^{1/2},
\end{align}
where $Z_q$ is the acoustic impedance of quartz, and $\eta$ ($\rho$) is the dynamic viscosity (mass density) of the liquid. The pseudo-differential technique requires matching the complex frequency response of the QCM to each liquid. This was done using the correct choice of liquids and experimental temperature. At $48^\circ$C, the growth of a water drop in undecane should result in $\Delta f_{drop}$ = $\Delta \Gamma_{drop}$ = 0 since $\eta_w\rho_w=\eta_u\rho_u$, where the subscripts refer to water (w) and undecane (u) throughout. Thus, deviations from zero reveal information about the differences in the liquid-solid boundary condition for each liquid. 

\begin{figure}
\begin{center}
\includegraphics[width=3.2 in]{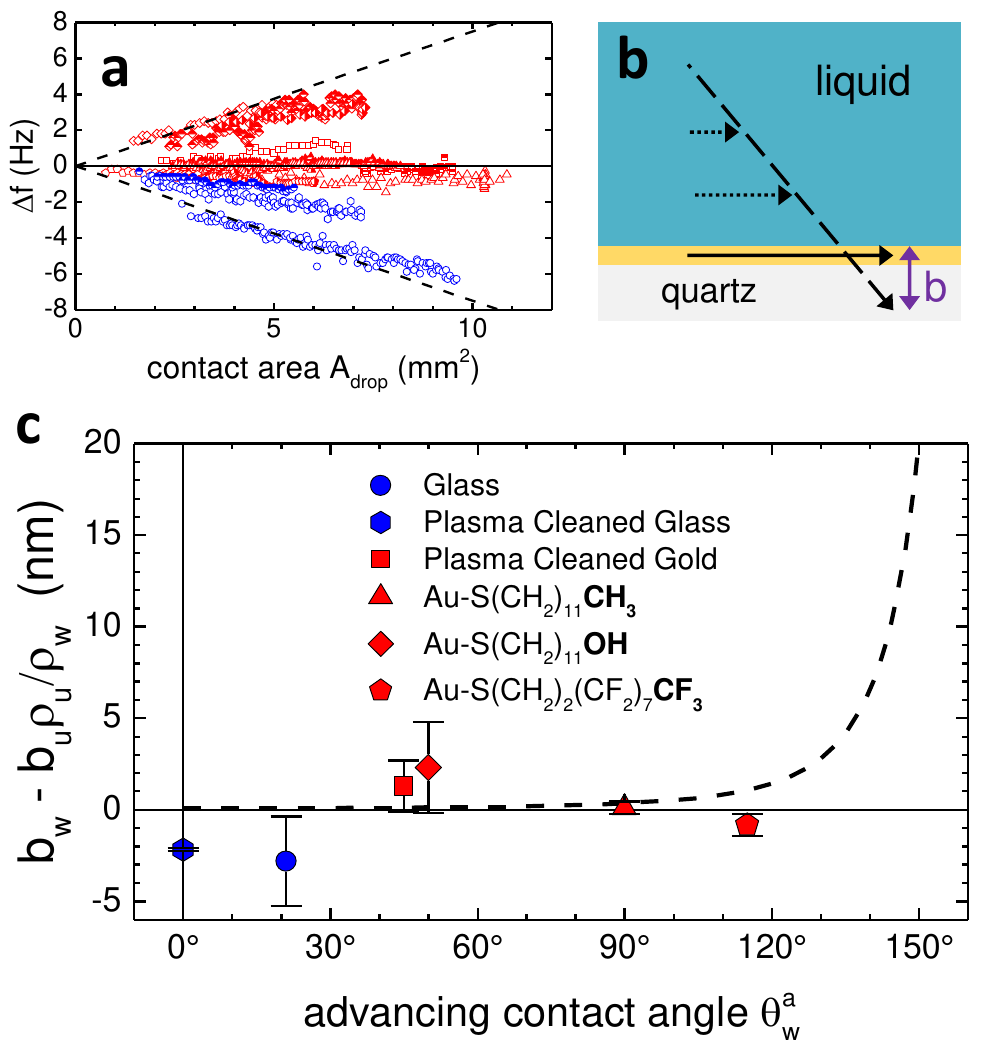}
\caption[]{(a) $\Delta f$ scales linearly with $A_{drop}$ on SiO$_2$ surfaces (blue), and gold and gold-SAM surfaces (red). (b) The slip length $b$ is defined by extrapolating the linear velocity profile at the surface. The dashed lines in (a) represent $b_w=\pm$ 2 nm, assuming $b_u$ = 0. (c) Differential slip length between water and undecane from Eqn.\ \ref{slipsig} versus advancing contact angle of the water. The dashed line is a theoretical prediction \cite{Huang2008}, as described in the text. Symbol shapes correspond to different surface chemistries, and symbol fills indicate different experiments for a given surface. Error bars represent variations from multiple experiments, linear fitting of the data, and optical contact area measurements.
\label{fig_2} }
\end{center} 
\end{figure}

Figure \ref{fig_2}a shows the shift in resonant frequency during the growth of water drops on various hydrophobic and hydrophilic surfaces. It is important to note that the signal grows linearly with contact area. If the signal was due to contact line effects, then the signal should scale with the perimeter of the drop ($\sim A_{drop}^{1/2}$). To leading order, slip is expected to decouple mass from the surface, leading to a change in frequency but not bandwidth \cite{Johannsmann2015,Huang2012,supp}:
\begin{align}
\label{slip1}
&\Delta f_{slip}=\dfrac{2 f_0^2 \rho}{Z_q}b,\\
\label{slip2}
&\Delta \Gamma_{slip}=\mathcal{O}(b^2/\delta^2),
\end{align}
where $\delta=\sqrt{\eta/\rho\pi f_0}$ is the viscous penetration depth associated with the decaying shear wave (Fig.\ \ref{fig_1}c), and is typically a few hundred nanometers for low-viscosity liquids such as water. 
If the liquid is slipping, then Eqn. \ref{slip1} predicts
\begin{align}
\label{slipsig}
&\Delta f_{drop}=2\dfrac{A_{drop}}{A_{active}}\dfrac{f_0^2 \rho_w}{Z_q}\left(b_w-\dfrac{\rho_u}{\rho_w}b_u\right)
\end{align}
upon the growth of a water drop with a contact area $A_{drop}$, where $A_{active}$ = 37.1 mm$^2$ is the measured active area of the QCM surface \cite{supp}. 

Figure \ref{fig_2}c shows the differential slip length from Eqn. \ref{slipsig} as a function of the advancing contact angle for a water drop in air, $\theta_w^a$. The data can also be interpreted as the absolute slip length of water on the various surfaces if we assume $b_u=0$. Naively, this is a reasonable assumption since undecane wets ($\theta^a_u=0^{\circ}$) all surfaces used in the experiment \cite{supp}. These results are consistent with recent simulations of water \cite{Huang2008,Sendner2009} indicating that slip only occurs for very hydrophobic surface interactions, although the shear rate used in the simulations ($10^{10}$ s$^{-1}$) was orders of magnitude larger than any experiment. The dashed line in Fig.\ \ref{fig_2}c represents a theoretical prediction for water \cite{Huang2008}, $b_w=0.5(1+\cos(\theta^a_w))^{-2}$ nm. The negative slip lengths consistently observed on SiO$_2$ surfaces may be related to the formation of solid-like water layers, which is enhanced by hydroxylation through oxygen-plasma cleaning \cite{Anderson2009} (Supplemental Fig.\ 4 \cite{supp}).

\begin{figure}
\begin{center}
\includegraphics[width=3.2 in]{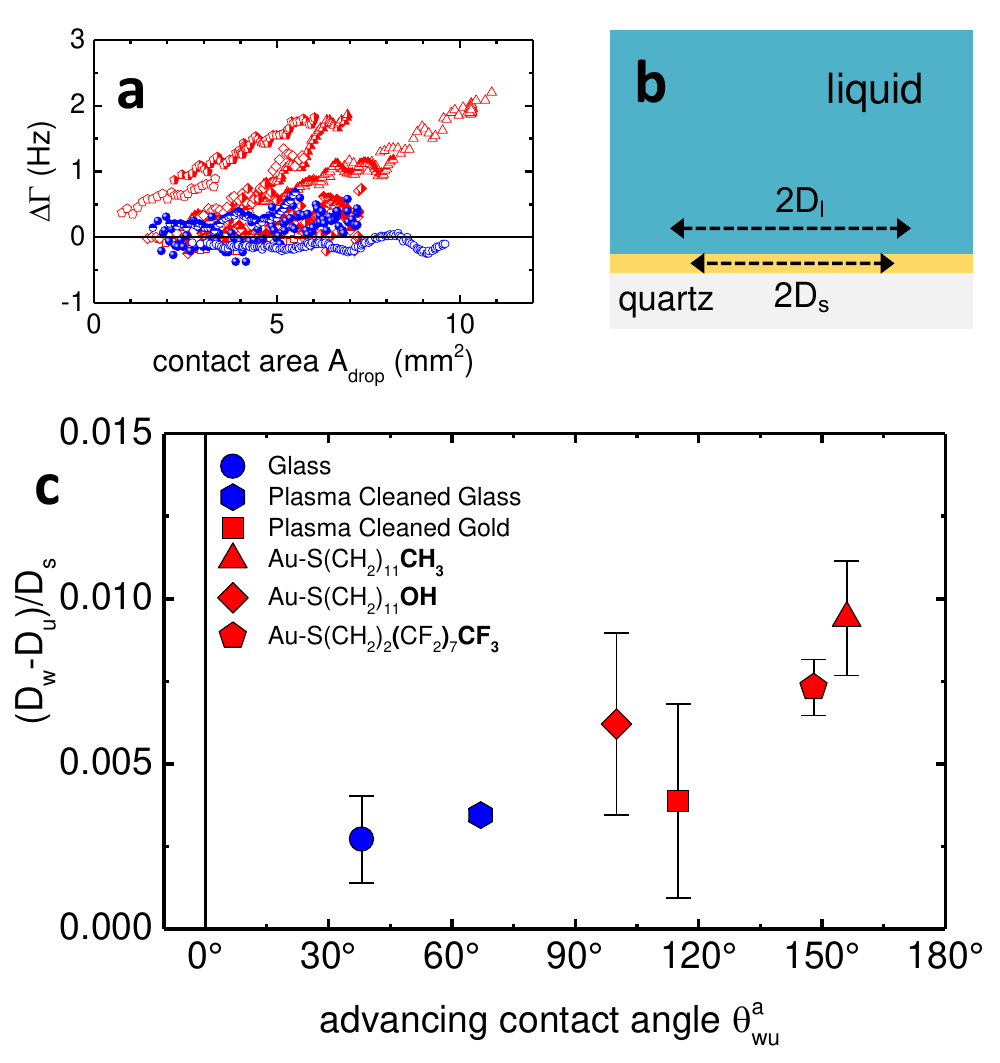}
\caption[]{(a) $\Delta\Gamma$ increases linearly with $A_{drop}$ on SiO$_2$ surfaces (blue), and on gold and gold-SAM surfaces (red). (b) Schematic showing a larger amplitude of motion in the liquid in comparison to the solid surface. (c) Difference in motional amplitude between the water and undecane at the solid surface, normalized by the amplitude of the solid, versus the advancing contact angle of the water drop in undecane. Symbol shapes correspond to different surface chemistries, and symbol fills indicate different experiments for a given surface. Error bars represent variations from multiple experiments, linear fitting of the data, and optical contact area measurements.
\label{fig_3}} 
\end{center} 
\end{figure}

Apparent slip can also be due to a thin surface layer of low density and viscosity \cite{Vinogradova1995,supp}, especially near hydrophobic surfaces where depleted, gas-like layers are expected to exist \cite{Doshi2005}. However, figure \ref{fig_2}c shows no obvious trend in the data for increased hydrophobicity. Conversely, Fig.\ \ref{fig_3}b shows an increase in $\Delta\Gamma$ as the water drop displaces the undecane, an effect which is larger for hydrophobic surfaces. 
The increase in dissipation is roughly linear in the contact area, and is not due to energy loss in the form of capillary waves, for example. We have verified this by reversing the role of the liquids in the experiment, i.e. growing an undecane drop on the surface in a water bath, where $\Delta\Gamma$ decreases by the same magnitude (Supplemental Fig.\ 3 \cite{supp}). An increase in $\Delta\Gamma$, with no corresponding change in $\Delta f$, is not easily explained. Variations in the liquid density or viscosity near the surface would primarily affect the resonant frequency of the QCM, not the bandwidth \cite{supp}. 

Nevertheless, an isolated change in $\Delta\Gamma$ can be explained by an additional elastic deformation between the liquid molecules and the SAM surface \cite{Johannsmann2015,Ferrante1994,supp}. For hydrophillic surfaces, the near-surface liquid molecules are sufficiently bound to the SAM layer so that they move synchronously with the QCM, and the amplitude of liquid motion is identical to the amplitude of solid motion. However, as the surface hydrophobicity increases, the near-surface molecules will experience excess deformation in their local potential minima, leading to a larger amplitude of motion in the liquid compared to the solid. Such deformation must occur prior on the onset of slip. Thus, upon the growth of a water drop, to leading order we expect \cite{supp}
\begin{align}
\label{film1}
&\Delta \Gamma_{drop}=2\dfrac{A_{drop}}{A_{active}}\left(\dfrac{D_w-D_u}{D_s}\right)\Delta\Gamma_{liq},
\end{align}
where $D_w$ ($D_u$) is the amplitude of motion of the water (undecane) at the surface, and $D_s$ is the amplitude of the solid, as shown in Fig.\ \ref{fig_3}b.  Figure \ref{fig_3}c shows the fractional increase in amplitude as a function of the advancing contact angle of the water drop in undecane, $\theta^a_{wu}$.  For the QCM voltages and liquids used in our experiment, the crystal amplitude is $D_s\approx$ 3.3 nm \cite{supp}, so the data in Fig.\ \ref{fig_3}c suggest the excess amplitude in the liquid is of order 0.1-0.3 \AA\, which is necessarily smaller than the typical lattice spacing for SAMs, given by $\sigma_{ss}\approx$ 5 \AA\  \cite{Barriet2003}. 

\begin{figure}
\begin{center}
\includegraphics[width=3.3 in]{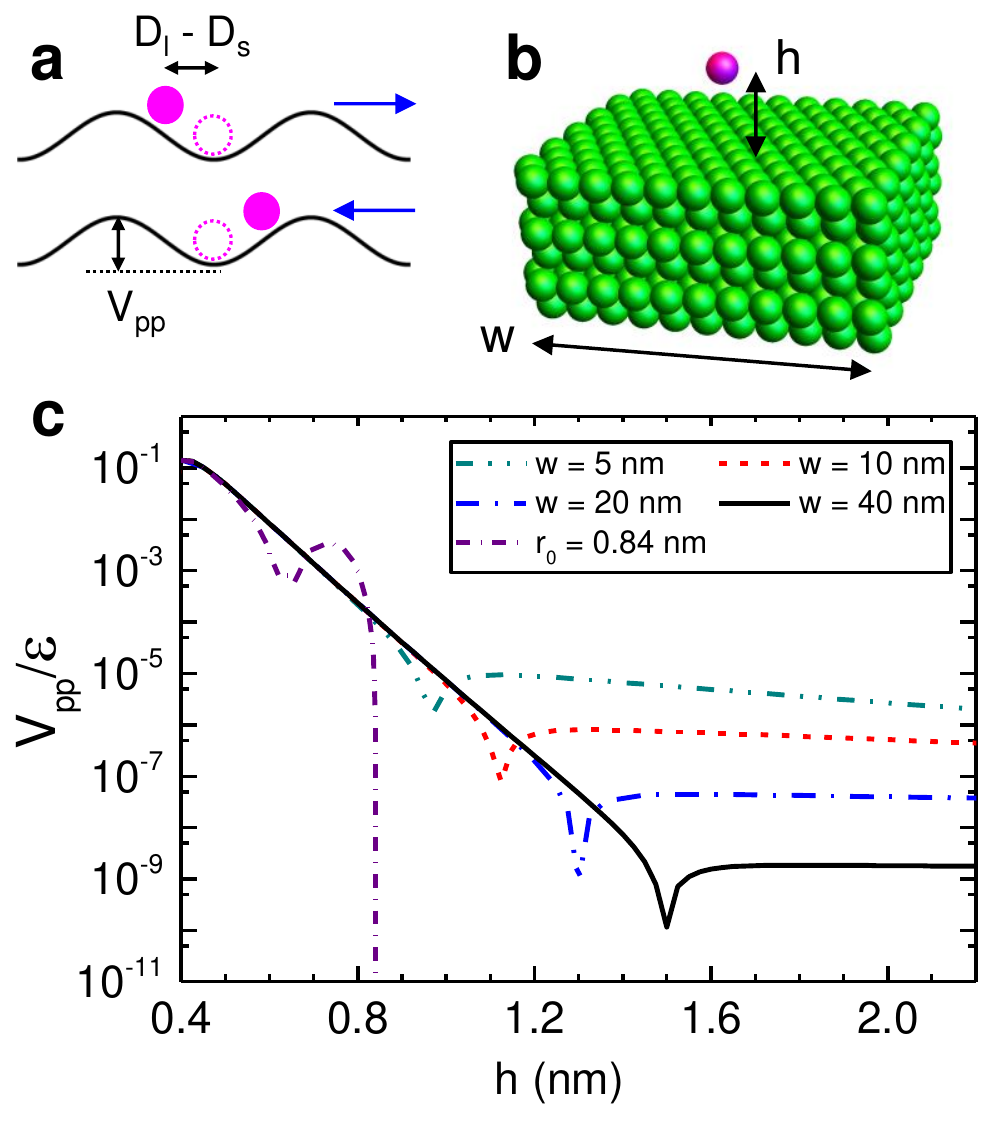}
\caption[]{(a) One-dimensional model showing a liquid molecule moving in the transverse potential during left and right oscillatory motion of the surface. The excess amplitude of the liquid is $D_l-D_s$. (b) Illustration of an fcc lattice of dimensions $w\times w\times w/2$ and a liquid molecule at a center-to-center distance $h$ above the top surface molecular plane. (c) Local peak-to-peak variation in potential, $V_{pp}$, normalized by the interaction energy, versus $h$. The local potential depends on the size of the lattice, as well as the cut-off radius $r_0$ used for the pair potential. Unless specified, $r_0=\infty$.
\label{fig_4}} 
\end{center} 
\end{figure}

This excess elastic deformation can not be provided by the SAM monolayer since the typical modulus of a SAM ($\sim$ 20 GPa \cite{Leng2000}) is orders of magnitude too large to produce the observed effect. Thus, deformation must be occurring at the liquid-SAM boundary. We can estimate the surface potential, $V_l$, experienced by a liquid molecule as the sum of all molecular pair-wise interactions across the interface. Near the surface, $V_l$ will have transverse, periodic variations due to the underlying lattice structure. The peak-to-peak amplitude of the variations, $V_{pp}$, will decay exponentially with surface separation \cite{Martini2008}. If the liquid is dragged along by the motion of the surface, then the stress provided by the transverse gradient in the potential must be sufficient to balance the shear stress in the liquid. This idea is illustrated in Fig.\ \ref{fig_4}a for a simple one-dimensional model. 

Locally, the potential is parabolic near a minimum, so the maximum stress provided by the transverse gradient is $(\hat{\bf t}\cdot\vec{\nabla}V_l)/\sigma_{ss}^2\approx 2\pi^2 V_{pp}(D_l-D_s)/\sigma_{ss}^4$, where $\hat{\bf t}$ is a unit tangent vector to the surface, and the subscript $l$ refers to the liquid phase. For the damped shear wave in the liquid, the maximum shear stress at the surface is $D_l(2\pi f_0)^{3/2}(\eta\rho)^{1/2}$ \cite{Johannsmann2015,supp}. Equating the two stresses gives
\begin{align}
\label{diss}
&\dfrac{D_l}{D_s}\approx 1+\dfrac{\sigma_{ss}^4}{V_{pp}}\left(\dfrac{2\eta\rho f_0^3}{\pi}\right)^{1/2},
\end{align}
where the approximation is valid when $D_l/D_s-1\ll 1$. Thus, for weak-binding substrates, the liquid will be, on average, further away from the surface, so $V_{pp}$ is smaller, leading to a larger compliance and an increase in the measured dissipation. 

In order to better understand how $V_{pp}$ depends on the distance from the surface, we compute the net potential from finite slabs of an fcc lattice of atoms with lattice constant $\sigma_{ss}$ = 5 \AA\ and dimension $w\times w\times w/2$, as shown in Fig.\ \ref{fig_4}b. The liquid-solid interaction is modeled using a standard 12-6 Lennard-Jones pair potential, $V(r)=4\epsilon((\sigma_{ls}/r)^{12}-(\sigma_{ls}/r)^6)+K_1 r+K_2$, where $r$ is the center-to-center separation and $\sigma_{ls}$ = 3.37 \AA\ is the length scale of the interaction between water and carbon \cite{Sendner2009}. The constants $K_1$ and $K_2$ are adjusted to satisfy $V(r_0)=0$ and $V'(r_0)=0$, where $r_0$ is the cut-off radius, set to infinity where noted. The total potential is found from summing over the particles in the lattice. Figure \ref{fig_4}c shows the sensitivity of $V_{pp}$ to the finite size of the system and range of the potential. Here, $V_{pp}$ is defined as the minimum potential barrier in any transverse direction along the surface. For $w\rightarrow\infty$, $V_{pp}$ exponentially decays away from the surface. For small slabs, the potential inverts ($V_{pp}\rightarrow 0$) before saturating. The largest effect is seen when the potential is cut off at $r_0$ = 0.84 nm \cite{Sendner2009}. 

Fig.\ \ref{fig_3}c shows that $(D_w-D_u)/D_s\approx 0.01$ for the most hydrophobic surfaces in our experiments. Assuming for simplicity that $D_u=D_s$ since $\theta_u^a=0$ for all surfaces, Eqn.\ \ref{diss} predicts that $V_{pp}\approx4.2\times 10^{-26}$ J for the water-solid interaction. For $\theta_w^a\sim120^{\circ}$, we can estimate a liquid-solid interaction energy, $\epsilon\approx5\times 10^{-22}$ J \cite{Sendner2009}. Thus $V_{pp}/\epsilon\approx 8.4\times 10^{-5}$, corresponding to a surface separation $z\approx 0.85$ nm in Fig.\ \ref{fig_4}c. The corresponding depletion length would be $h-\sigma_{ls}$ = 5.1 \AA\ , which is in excellent agreement with both neutron \cite{Doshi2005} and x-ray reflectivity measurements  \cite{Chattopadhyay2010} of depletion layers for SAM-water interfaces, although some controversy exists on the latter \cite{Mezger2011,Chattopadhyay2011}.

We expect slip to occur by increasing the hydrophobicity (decrease $V_{pp}$), or by increasing the shear stress (increase $\dot{\gamma}$). The maximum gradient in the potential occurs at 1/4th the lattice spacing from the local minimum, as shown in Fig. \ref{fig_4}a. This corresponds to $(D_l-D_s)/D_s=\sigma_{ss}/4D_s\approx$ 0.04. Thus, increasing the shear rate by a factor of 4 would exceed the maximum tangential force provided by the surface, resulting in slip for $\theta_{w}^a\approx$ 120$^{\circ}$. This would correspond to $\dot{\gamma}\approx 3.1\times 10^6$ s$^{-1}$, which is a factor of 100 less than the smallest shear rates in molecular dynamics simulations where slip is often reported. 

Nevertheless, several experimental studies report slip lengths of order $b\approx10-100$ nm for water on hydrophobic surfaces composed of smooth glass coated with a silane-based SAM \cite{Bocquet2010,Vinogradova2011,Cottin-Bizonne2005,McBride2009,Vinogradova2009,Joly2006}, as opposed to the gold-thiol based surfaces reported here. Although a more recent study with higher experimental sensitivity show data consistent with $b=0$ nm on silanized surfaces \cite{Schaeffel2013}, our results highlight that equilibrium contact angle alone is not sufficient to characterize the slip behavior of liquids near smooth surfaces. Of equal importance is the shear rate threshold for slip \cite{Martini2008}, which is directly related to transverse variations in the surface potential, and can not be captured by continuum models where slip is confined to a near-surface layer \cite{Vinogradova1995}. For silane-based SAMs, we hypothesize that the consistently low degree of translational ordering \cite{Schreiber2000} may lead to a broad distribution of energy barriers in the corrugated surface potential, so that slip may be induced at low shear rates, however, such a hypothesis would need to verified.

In summary, we have demonstrated that precursors to slip can be observed for hydrophobic surfaces at large experimental shear rates, a result which connects experiments and simulations that are necessarily separated by orders of magnitude in dynamical time scales, and will impact multiple fields of research involving nanoscale liquid dynamics near material and biological interfaces. In addition, although the range of the interaction potential in simulations has been considered for static quantities such as the equilibrium contact angle \cite{Sendner2009}, we have shown here that it is equally important for the interfacial dynamics. As applications necessitate even smaller length scales for flows, confinement effects become more important, as well as increased shear rates, both of which remain open questions for ongoing research efforts. 

\begin{acknowledgments} 
This work was supported by the NSF DMR Grant No. 1455086. We thank H. Stone, D. Lohse, and F. Family for fruitful discussions.
\end{acknowledgments} 

%


\end{document}